\documentclass{raa}            
\usepackage{graphicx,times}             
\usepackage{natbib}
\usepackage{amssymb,amsmath}
\bibpunct{(}{)}{;}{a}{}{,}
\usepackage[pagebackref=true]{hyperref}
\usepackage{siunitx}

\newcommand{\Rmnum}[1]{\expandafter\@slowromancap\romannumeral #1@}

\begin{document}

  \title{Cosmic Dynamics in Einstein-Cartan Theory: Analysing Hubble Tension through Curvature and linear Torsion field}
   \volnopage{Vol.0 (20xx) No.0, 000--000}      
   \setcounter{page}{1}          

   \author{Yun-Dong Wu 
      \inst{1,2}
   \and Wei Hong
      \inst{1,2}
   \and Tong-Jie Zhang
      \inst{1,2}
   }

\institute{Institute for Frontiers in Astronomy and Astrophysics, Beijing Normal University, Beijing 102206, China; {\it tjzhang@bnu.edu.cn}\\
        \and
             School of Physics and Astronomy, Beijing Normal University, Beijing 100875, China.\\
\vs\no
   {\small Received 2025 Oct 18; revised 2026 Feb 11; accepted 2026 Feb 26}}

\abstract{ The Hubble tension refers to the significant discrepancy in the Hubble constant $H_{0}$ obtained from two different measurement methods in cosmology which has persisted for decades. To theoretically explore potential solutions to this problem, this paper examines a model within the framework of Einstein-Cartan (EC) theory, where torsion is introduced with spin as the corresponding entity, allowing for a linear assumption between $H$ and $\phi$. By employing the Markov Chain Monte Carlo (MCMC) algorithm and utilizing Cosmic Chronometers (CC) data, we impose parameter constraints on various parameters in the Friedmann equations, particularly focusing on the curvature density parameter $\Omega_k$, to assess whether the model remains stable under this assumption and whether the estimated parameters align more closely with either of the observational results. In conclusion, we find that the parameter constraints in the model incorporating torsion ($ H_0 = 67.6^{+2.1}_{-2.7} \ \mathrm{km\ s^{-1}\ Mpc^{-1}}$, obtained under the Big Bang Nucleosynthesis (BBN) constraint with $\Omega_{k}=0$; $ H_0 = 66.2^{+4.4}_{-2.9} \ \mathrm{km\ s^{-1}\ Mpc^{-1}}$, obtained under same constraint but set $\Omega_{k}$ as a free variable; $ H_0 = 68.8^{+2.9}_{-4.2} \ \mathrm{km\ s^{-1}\ Mpc^{-1}}$, obtained under the Planck constraint) are more consistent with the value derived from CMB data, favoring lower $H_0$ value.
}

   \authorrunning{Y.-D. Wu, W. Hong \& T.-J. Zhang }          
   \titlerunning{Analyzing Hubble Tension via Curvature and Torsion}  

   \maketitle

\section{Introduction}
Since Einstein proposed General Relativity in 1915, the theory has successfully explained a range of astronomical phenomena, including the precession of Mercury's perihelion and gravitational lensing, becoming the cornerstone of modern gravitational theory (\citealt{1915SPAW.......844E}). However, this framework is not complete in some aspects. One of its limitations is the inability to account for the spacetime structure in the presence of fermions with spin angular momentum. To address this, Elie Cartan introduced torsion in 1922 as the spin counterpart in the \cite{Cartan:1923zea}, partially filling the gap in General Relativity's description of the impact of spinning particles on spacetime. 

By the 1950s, scholars had widely integrated General Relativity with spin theory, gradually forming the Einstein-Cartan (EC) theory (\citealt{1976RvMP...48..393H}), which successfully described certain observational phenomena (albeit not uniquely). Examples include using EC theory to describe massive, dense objects (\citealt{1974PhRvD..10.1066H, Jockel2024fps}) and interpreting dark energy as torsion to explain the dynamical evolution of the universe (\citealt{AO2010186}). To this day, the theory continues to evolve and serves as the foundation for the discussions in this paper (\citealt{torsion_field_equations, Benisty_2022}). 

In EC theory, the Riemann curvature tensor can be written as: $ R^\rho_{\sigma\mu\nu} = \partial_\mu \Gamma^\rho_{\nu\sigma} - \partial_\nu \Gamma^\rho_{\mu\sigma} + \Gamma^\rho_{\mu\lambda} \Gamma^\lambda_{\nu\sigma} - \Gamma^\rho_{\nu\lambda} \Gamma^\lambda_{\mu\sigma} $. In the case of symmetric Christoffel symbols (under the torsion-free condition), the curvature tensor also exhibits symmetry, reflecting the symmetric and global structure of the spacetime. In contrast, the torsion tensor behaves differently. It is defined as $ S^{\rho}_{\mu \nu} = \Gamma^{\rho}_{[\mu \nu]} $, and is an anti-symmetric tensor. Moreover, it is not difficult to see that its definition does not specify a zero point, which implies that torsion is always a relative value, reflecting the degree of deviation from the symmetric structure. However, when we contract the Riemann tensor to obtain the Ricci scalar, $ \tilde{R} = R + K_{\mu \nu \rho} K^{\mu \nu \rho} $, we find that compared to the tensor contraction result $R$ of torsion-free case, an additional summation term involving the contorsion tensor appears. This term stands alongside the spatial curvature term, $ \frac{k}{a^{2}} $, in the original Ricci scalar. This indicates that although curvature and torsion influence the geometry of spacetime in different ways, their ultimate impact on the dynamical factors is similar.

There is currently no direct observational evidence for torsion, as its associated physical quantities require observations in extreme environments, such as high matter or energy densities. Nevertheless, this does not diminish the potential of EC theory as a promising framework for explaining the universe. Over the past decade, EC theory has achieved significant advancements in a variety of research directions. It has been applied, for instance, to addressing the strong CP problem (\citealt{2024JHEP...11..146K}), explaining the intrinsic mechanisms behind the accelerated expansion of the universe (\citealt{2018EPJC...78...38H}), and emulating the evolution of dark Energy and dark Matter (\citealt{2025PDU....4801848M}). It also plays a role in large-scale structural issues, where, combined with numerical simulation methods, it has been used to explore the impact of torsion-induced density perturbations on the evolution of cosmic structures, including large-scale formations (\citealt{2023EPJC...83..958U}) and cosmic strings (\citealt{2006FizB...15....1R}) in the universe.

In other areas, significant progress has been made in EC theory. Efforts have been made to maintain Weyl invariance in both global and local formulations (\citealt{2021PhRvD.104l4014K, Gialamas_2025}), while the homogeneous spin fluid model has been utilized to address the theoretical issue of singularity formation (\citealt{2015EPJC...75...53H}). The theory has also been expanded into the Einstein-Cartan-Kibble-Sciama (ECKS) theory, which introduces torsion to generate a strong repulsive force at small scales, thus preventing the formation of singularities (\citealt{2010PhLB..694..181P}). This approach offers an alternative to the classical inflation model in explaining the early universe’s expansion. Notably, the oscillation model of the Higgs particle within EC theory also stands as a strong candidate for explaining inflation (\citealt{Piani_2023}).
Among the various research directions, gravitational waves have become a particularly active field. In \cite{2022EPJC...82..628B}, the authors recalculated the Weyssenhoff-fluid model under torsion conditions, deriving the power of gravitational waves emitted by binary neutron stars in the post-Newtonian approximation. Similar studies (\citealt{2023PDU....4001197E, 2024EPJC...84..316R}) have further explored gravitational waves in the context of EC theory. 
These series of attempts to integrate popular research topics with EC theory directly reflects the necessity of exploring its combination with EC theory in addressing the Hubble tension problem (\citealt{Izaurieta_2020}). 

The analysis in this paper employs parameter constraints to explore the dynamic evolution of torsion and curvature in the universe, representing a further study of EC theory under the FLRW metric. The main research object, torsion tensor, is a (1,2)-type tensor field on an n-dimensional affine connection space (M, D), while the curvature tensor is a (1,3)-type tensor field. Both of them are important invariants that characterize the differences between an affine connection space and an affine space. Therefore, the fundamental objective of our research is to provide an explanation for the asymmetrical phenomena observed in the universe, with the aspiration that its ultimate findings may offer a framework for addressing the Hubble tension.

\section{Theoretical foundation}

\subsection{Spacetime with torsion}
 In Einstein Cartan theory, torsion tensor is defined by anti-symmetric part of the affine connection, as presented in $S_{bc}^{a}=\Gamma_{[bc]}^{a}$. As we always do, demanding metric tensor satisfy the covariant condition as $\nabla_{c}g_{ab}=0$. This result is oriented towards the following equation of connection:
 \begin{equation}
     \Gamma_{\ bc}^{a} = \tilde{\Gamma}_{\ bc}^{a} + K_{\ bc}^{a}.
     \label{eq1}
 \end{equation}
 First part gives typical Christoffel symbols, which defined in detail in Appendix A. And the second part is contortion tensor which defined as below:
\begin{equation}
    K_{\ bc}^{a} = S_{abc} + S_{bca} + S_{cba} = S_{abc} + 2S_{(bc)a}, 
    \label{eq2}
\end{equation}
where $S_{abc}$ is torsion tensor. The equation $K_{abc}=K_{a[bc]}$ leads to an interesting phenomenon: the transformation of infinitesimal parallelograms is non-closing. From a physical perspective, this implies that the decomposition can effectively characterize torsion, which is typically considered a manifestation of the endowed angular momentum of particle.
The anti-symmetry of $S_{abc}$ guarantees it has only one independent tensor contraction, so we can make an intuitive definition of anti-symmetric part of contortion tensor with its' contraction:

\begin{equation}
    S_{a} = S_{\ ab}^{b} = -S_{\ ba}^{b} \equiv \phi, \quad a=0,1,2,3.
    \label{torsion scalar}
\end{equation})

\subsection{Field equations with torsion}
 In non-torsion spacetime, according to GR theory we can directly write the formula:
\begin{equation}
    R_{ab} - \frac{1}{2}g_{ab}R = \kappa T_{ab} - \Lambda g_{ab},
\end{equation}
where $R_{ab}$ represents the Ricci tensor and $R = R_{\ a}^{a}$. This field equation follows the standard form. Similarly, in Einstein-Cartan theory, the field equations may retain this form, but the presence of torsion alters the expression of the Ricci tensor. In this theory, $R_{ab}$ and $T_{ab}$ become anti-symmetric tensors. In Cartan field equations, the torsion tensor is typically defined as:
\begin{equation}
    S_{abc} = -\frac{1}{4}\kappa (2s_{bca} + g_{ca}s_{b} - g_{ab}s_{c}),
\end{equation}
where $\kappa = 8\pi G$ and $s_{abc}$ represents the spin tensor of matter with $s_{abc} = s_{[ab]c} $ , $s_{a} = s_{\ ab}^{b}$. To satisfy the Bianchi identities, we can derive the following relationship between the Ricci tensor and the torsion tensor under this condition through a series of calculations:
\begin{equation}
    \nabla_{[e}R^{ab}_{\ \ cd]} = 2R^{ab}_{\ \ f[e}S_{\ cd]}^{f},
\end{equation}
and
\begin{equation}
    R_{\ [bcd]}^{a} = -2\nabla_{[b}S_{\ cd]}^{a} + 4S_{\ e[b}^{a}S_{\ cd]}^{e}.
\end{equation}
Next, by formulating a similar definition to the former $G_{ab} = R_{ab} - \frac{1}{2}g_{ab}R$, we can derive the final corrected covariant derivative (This result has been proved in \cite{torsion_field_equations} and we will find that it does not satisfy the general relativistic conservation law $\nabla^{b}G_{ab} = 0$.
\begin{equation}
    \nabla^{b}G_{ab} = \nabla^{b}G_{ba} - 2(\nabla^{2}S_{a} - \nabla^{b}\nabla_{a}S_{b} + \nabla^{b}\nabla^{c}S_{cba}) - 4\nabla^{b}(S^{c}S_{cab}).
\end{equation}

\subsection{Friedmann equation with torsion in FLRW metric}
 We first consider the metric associated with torsion. Using the expression 
$\Gamma_{\ bc}^{a} = \tilde{\Gamma}_{\ (bc)}^{a} + 2S_{bc}^{\quad a}$ (which derived from Eqs. \eqref{eq1}, \eqref{eq2}) and its transformation, we can compute the Christoffel symbols with torsion from the torsion-free case. 
Due to the complexity of the specific metric calculations, the results are provided in detail in the appendix at the end of the paper. 
 
 Similarly, we can still calculate the Ricci tensor using the following definition:
\begin{equation}
    R_{ab} = -\partial_{b}\Gamma_{\ ac}^{c} + \partial_{c}\Gamma_{\ ab}^{c} - \Gamma_{\ ac}^{e}\Gamma_{\ eb}^{c} + \Gamma_{\ ab}^{e}\Gamma_{\ ec}^{c} .
\end{equation}
Then, we assume that the energy-momentum tensor can be described by an perfect fluid model,
\begin{equation}
    T_{ab} = \rho u_{a}u_{b} + p\eta_{ab},
\end{equation}
where $u_{a}u^{a} = -1$ is time-like 4-velocity and $\eta_{ab}$ is Minkowski metric. Given the aforementioned conditions, we ultimately achieve the Friedmann equation:
\begin{equation}
    (\frac{\dot{a}}{a})^{2} = \frac{1}{3}\kappa\rho - \frac{K}{a^{2}} + \frac{\Lambda}{3} - 4\phi^{2} - 4(\frac{\dot{a}}{a})\phi,
\end{equation}
and acceleration equation with torsion:
\begin{equation}
    \frac{\Ddot{a}}{a} = -\frac{1}{6}\kappa(\rho + 3p) + \frac{\Lambda}{3} -2\dot{\phi} - 2(\frac{\dot{a}}{a})\phi.
    \label{acceleration equation}
\end{equation}
Further, equation of state can be written as $\rho = \omega p $, where $ \rho$ is energy density, $p$ is pressure and $\omega$ is state parameter. Then we can give the continuity equation by $ \nabla_{a}T^{ab} = 0$ as below:
\begin{equation}
    \dot{\rho} + 3(1+\omega)H\rho + 2(1+3\omega)\phi \rho = 4\phi \frac{\Lambda}{8\pi G}.
    \label{continuity equation}
\end{equation}

\subsection{Hubble parameter's behavior under linear dependence to torsion field}
In Eq. \eqref{torsion scalar}, we defined that anti-symmetric part of torsion tensor only have one independent contraction. Therefore, we can give a further hypothesize its specific form of association with other parameters from \cite{torsion_form}. Here, we focus on a special form: $ \phi(t) = -\alpha H(t)$ where $H = \frac{\dot{a}}{a}$ is Hubble parameter, $a$ is scale factor of universe and $\alpha $ is a constant which represents the strength of torsion field. 

Now, we can analyze the Friedmann equations under above conditions. The first step is to dimensionless the continuity equation. Let $ \rho = x\rho_{0}$, where $x$ is dimensionless energy density and $\rho_{0}$ is the energy density $ \rho $ in the situation of $a=1$. Then we can turn Eq. \eqref{continuity equation} into:
\begin{equation}
    \dot{x} + 3(1+\omega)Hx + 2(1+3\omega)\phi x = 4\phi\frac{\Lambda}{8\pi G\rho_{0}} .
    \label{dimensionless}
\end{equation}
 To simplify the calculation, we now define the present density parameters $ \Omega_{m0} = \frac{8\pi G\rho_{m0}}{3H_{0}^{2}}$ and $ \Omega_{\Lambda0} = \frac{\Lambda}{3H_{0}^{2}}$, then the above equation change to:
 \begin{equation}
     \frac{dx}{da} + [3(1+\omega)-2\alpha(1+3\omega)]\frac{x}{a} + 4\alpha\frac{\Omega_{\Lambda0}}{\Omega_{m0}} = 0.
 \end{equation}
 Integrating after separating the parameters gives:
 \begin{equation}
     x = a^{-C_{1}}(1+\frac{C_{2}}{C_{1}}) - \frac{C_{2}}{C_{1}},
     \label{r(a)}
 \end{equation}
 where $ C_{1} = (3-2\alpha) + 3\omega(1-2\alpha)$ and $ C_{2} = 4\alpha\frac{\Omega_{\Lambda0}}{\Omega_{m0}}$.
  
Second, bringing the above assumption into acceleration equation, Eq. \eqref{acceleration equation}, it turns to following form:
\begin{equation}
    (1-2\alpha)\frac{\dot{a}}{a}\frac{d\dot{a}}{da} = H_{0}^{2}[\Omega_{\Lambda0} - \frac{(1+3\omega)}{2}\Omega_{m0}x].
    \label{developed acceleration eq}
\end{equation}
Substitute the Eq. \eqref{r(a)} into Eq. \eqref{developed acceleration eq} and integrate by separating parameters. Subsequently, we can give the expression of the Hubble parameter in terms of the parameters which we aim to constrain:
\begin{equation}
    H_{1}(z) = (C_{3} - C_{4}(1+z)^{C_{1}} + C_{5}(1+z)^{2})^{\frac{1}{2}},
    \label{h1}
\end{equation}
where $ C_{3} = \frac{H_{0}^{2}}{1-2\alpha}(\Omega_{\Lambda0}+\frac{1+3\omega}{2}\frac{C_{2}}{C_{1}}\Omega_{m0})$, $ C_{4} = \frac{H_{0}^{2}}{1-2\alpha}(1+3\omega)(1+\frac{C_{2}}{C_{1}})\frac{\Omega_{m0}}{2-C_{1}} $, $ C_{5} = C_{4} - C_{3} + H_{0}^{2}$ containing variable parameter $H_{0},\Omega_{\Lambda0},\Omega_{m0}, \omega \ and\  \alpha $. However, the curvature density parameter is of primary interest in this study and is not included among these parameters. Therefore, we need to incorporate the simplified form of the first Friedman Eq. \eqref{dimensionless} into the final expression as well:
\begin{equation}
    H_{2}(z) = \frac{H_{0}}{|1-2\alpha|}[((1+z)^{C_{1}}(1+\frac{C_{2}}{C_{1}}) - \frac{C_{2}}{C_{1}})\Omega_{m0} + \Omega_{k}(1+z)^{2} + \Omega_{\Lambda}]^{\frac{1}{2}}.
    \label{h2}
\end{equation} 
In the final numerical simulations, both formulas are components of the likelihood function.

\section{Numerical methods}
Now, we use the previously computed formulas Eqs. \eqref{h1} and \eqref{h2} to constrain these parameters. The parameter constraint method used here is a common approach in cosmology for performing analysis on data, which continues to be updated and developed to this day (\citealt{2023ApJS..268...67H}). The reason for utilizing both formulas simultaneously in the simulation is that, in essence, the two formulas are equivalent. And theoretically, starting from either should yield to the same result. However, the former does not explicitly include the $\Omega_{k}$, which we require as a variable in this context. Additionally, to maintain the stability of the MCMC process, we incorporate both formulas in the simulation.
In our simulations, the variables are the Hubble parameter $H$ and red-shift $ z$, data sourced from \cite{CC}, referred to as CC. At the end of this section, Table. \ref{CC data} is presented, compiling the specific CC reference data sources cited earlier in the paper. This study employs the MCMC method for data handling, which combining Monte Carlo methods with Markov chains. 

The aforementioned method is employed here because the sample size of CC is very small. We must increase the sample size, and the MCMC method is particularly well-suited for this purpose as it can generate samples from the target distribution to meet our simulation requirements. For convenience, a non-informative prior distribution is used here. This also indicates that we do not overly rely on any specific assumptions or expectations during the model construction process. The next step involves specifying the likelihood function. Assuming the likelihood function to be normally distributed without loss of generality, then by combining the two formulas provided above, we can derive the specific expression of the likelihood function with the CC data:
\begin{equation}
    ln(L_{H}) =\sum_{j=1,2}^{Eqs. \eqref{h1},\eqref{h2}}(\sum_{i=1}^{n}{-\frac{[H_{obs,i}-H_{j}(z)]^{2}}{2\sigma_{H_{obs,i}}^{2}} -\frac{1}{2}ln(2\pi\sigma_{H_{obs,i}}^{2} }) ),
\end{equation}
where $ln(L_{H})$ is log-likelihood function, $H_{obs}, \sigma_{H_{obs}}$ represent the actual observed values and the standard deviation of the observed values of the Hubble parameter in CC and H(z) represents the theoretical predicted value of the Hubble parameter at the current red-shift.

\begin{table}[ht!]
\footnotesize
\caption{Measurements of the Hubble parameter were derived with the Cosmic Chronometers method in units of $\mathrm{km\ s^{-1}} \mathrm{Mpc^{-1}}$. In this table, “M" represents the method used to derive the Hubble parameter. F (Full-Spectrum Fitting) fits the entire galaxy spectrum to estimate stellar ages (\citealt{https://doi.org/10.1046/j.1365-8711.2003.06897.x}), L (Lick Indices) analyzes specific absorption lines to determine age (\citealt{1994ApJS...94..687W}), D (D4000) measures the 4000 Å break to estimate stellar age (\citealt{1999ApJ...527...54B}), and ML (Machine Learning) applies algorithms to large datasets for estimating stellar ages and cosmological parameters (\citealt{2021ApJ...915...71V}).}
\label{CC data}
\doublerulesep 0.1pt \tabcolsep 13pt
\centering
\begin{tabular}{ccccc}
\hline
\textbf{z} & \textbf{H(z)} & \textbf{$\sigma_{H(z)}$ } & \textbf{M} & \textbf{Reference}\\
\hline
0.07  & 69.0  & 19.6  & F  & \cite{2014RAA....14.1221Z} \\ 
0.09  & 69    & 12    & F  & \cite{2005PhRvD..71l3001S} \\
0.12  & 68.6  & 26.2  & F  & \cite{2014RAA....14.1221Z} \\ 
0.17  & 83    & 8     & F  & \cite{2005PhRvD..71l3001S} \\ 
0.179 & 75    & 4     & D  & \cite{2012JCAP...08..006M} \\
0.199 & 75    & 5     & D  & \cite{2012JCAP...08..006M} \\ 
0.20  & 72.9  & 29.6  & F  & \cite{2014RAA....14.1221Z} \\ 
0.27  & 77    & 14    & F  & \cite{2005PhRvD..71l3001S} \\ 
0.28  & 88.8  & 36.6  & F  & \cite{2014RAA....14.1221Z} \\
0.352 & 83    & 14    & D  & \cite{2012JCAP...08..006M} \\ 
0.38  & 83    & 13.5  & D  & \cite{2016JCAP...05..014M} \\ 
0.4   & 95    & 17    & F  & \cite{2005PhRvD..71l3001S} \\ 
0.4004 & 77   & 10.2  & D  & \cite{2016JCAP...05..014M} \\ 
0.425  & 87.1 & 11.2  & D  & \cite{2016JCAP...05..014M} \\ 
0.445  & 92.8 & 12.9  & D  & \cite{2016JCAP...05..014M} \\ 
0.47   & 89.0 & 49.6  & F  & \cite{2017MNRAS.467.3239R} \\ 
0.4783 & 80.9 & 9     & D  & \cite{2016JCAP...05..014M} \\ 
0.48   & 97   & 62    & F  & \cite{2010JCAP...02..008S} \\
0.593  & 104   & 13    & D  & \cite{2012JCAP...08..006M}\\
0.68   & 92    & 8     & D  & \cite{2012JCAP...08..006M}\\
0.75   & 98.8  & 33.6  & L  & \cite{2022ApJ...928L...4B}\\
0.75   & 105   & 10.76 & ML & \cite{2023JCAP...11..047J}\\
0.781  & 105   & 12    & D  & \cite{2012JCAP...08..006M}\\
0.8    & 113.1 & 25.22 & F  & \cite{2023ApJS..265...48J}\\
0.875  & 125   & 17    & D  & \cite{2012JCAP...08..006M}\\
0.88   & 90    & 40    & F  & \cite{2010JCAP...02..008S}\\
0.9    & 117   & 23    & F  & \cite{2005PhRvD..71l3001S}\\
1.037  & 154   & 20    & D  & \cite{2012JCAP...08..006M}\\
1.26   & 135   & 65    & F  & \cite{2023AA...679A..96T} \\
1.3    & 168   & 17    & F  & \cite{2005PhRvD..71l3001S}\\
1.363  & 160   & 33.6  & D  & \cite{2015MNRAS.450L..16M}\\
1.43   & 177   & 18    & F  & \cite{2005PhRvD..71l3001S}\\
1.53   & 140   & 14    & F  & \cite{2005PhRvD..71l3001S}\\
1.75   & 202   & 40    & F  & \cite{2005PhRvD..71l3001S}\\
1.965  & 186.5 & 50.4  & D  & \cite{2015MNRAS.450L..16M}\\

\hline
\end{tabular}
\end{table}

The parameters constraint in this study are performed using Python, utilizing the MCMC computational capabilities of the Emcee package in \cite{emcee}. Additionally, it should be noted that the priors in this study include direct constraints on the curvature density parameter. The specific constraint range has been fine-tuned through a series of adjustments, but is initially based on observational data from \cite{Plank2018}. Now presenting the approximate prior range used in the program: 1. Planck condition: A. $0 < H < 100\ \mathrm{km\ s^{-1}} \mathrm{Mpc^{-1}}$; B. $ -0.1 < \alpha < 0.1$; C. $- 0.0012 < \Omega_{k} < 0.0026$. 2. BBN condition: $H$ and $\alpha$ limitation are the same as Planck condition, except C. $ 0.02202 < \Omega_{m}(H/100)^{2} < 0.02268 $. There are two reasons for applying different priors in this context. First, these various priors effectively integrate constraints from a range of observational data and theoretical models, and testing the stability of the model under these constraints is, in itself, a way to evaluate the validity of the theory. Second, the use of direct constraints ensures that the parameter estimates remain close to the stable value, rather than near the metastable state.

\section{Results Analysis}
\subsection{Information from figures}

\begin{table}[ht!]
\centering
\footnotesize
\caption{The unit of Hubble Parameter is same as Table.\ref{CC data} ($\mathrm{km\ s^{-1}} \mathrm{Mpc^{-1}}$). The purpose of this table is to present the necessary data from the figures of this section in a concentrated format.}

\label{Hubble Data}
\centering
\begin{tabular}{cccccc}
\hline
\textbf{Figure} & \textbf{$H$} & \textbf{$\Omega_{\Lambda}$ } & \textbf{$\Omega_{m}$} & \textbf{$\Omega_{k}$} & \textbf{$\alpha$}\\
\hline
\ref{ms2025-0469fig1} & $67.9_{-2.1}^{+2.3}$ & $0.681_{-0.042}^{+0.042}$ & $0.316_{-0.039}^{+0.045}$ & / & / \\
\ref{ms2025-0469fig2}& $67.6_{-2.2}^{+2.7}$ & $0.681_{-0.047}^{+0.037}$ & $0.320_{-0.038}^{+0.046}$ & $0.001_{-0.054}^{+0.053}$ & / \\
\ref{ms2025-0469fig3} & $67.6_{-2.7}^{+2.1}$ & $0.556_{-0.067}^{+0.062}$ & $0.246_{-0.029}^{+0.036}$ & / & $0.0517_{-0.0071}^{+0.0094}$ \\
\ref{ms2025-0469fig4} & $66.2_{-2.9}^{+4.4}$ & $0.60_{-0.28}^{+0.21}$ & $0.277_{-0.120}^{+0.087}$ & $ -0.13_{-0.36}^{+0.54}$ & $0.061_{-0.0036}^{+0.0024}$ \\
\ref{ms2025-0469fig5} & $68.8_{-4.2}^{+2.9}$ & $0.75_{-0.34}^{+0.36}$ & $0.36_{-0.19}^{+0.19}$ & $ -0.07_{-0.46}^{+0.49}$ & $-0.001_{-0.101}^{+0.065}$ \\

\hline
\end{tabular}
\end{table}

\begin{figure}
\centering
\includegraphics[width=\linewidth]{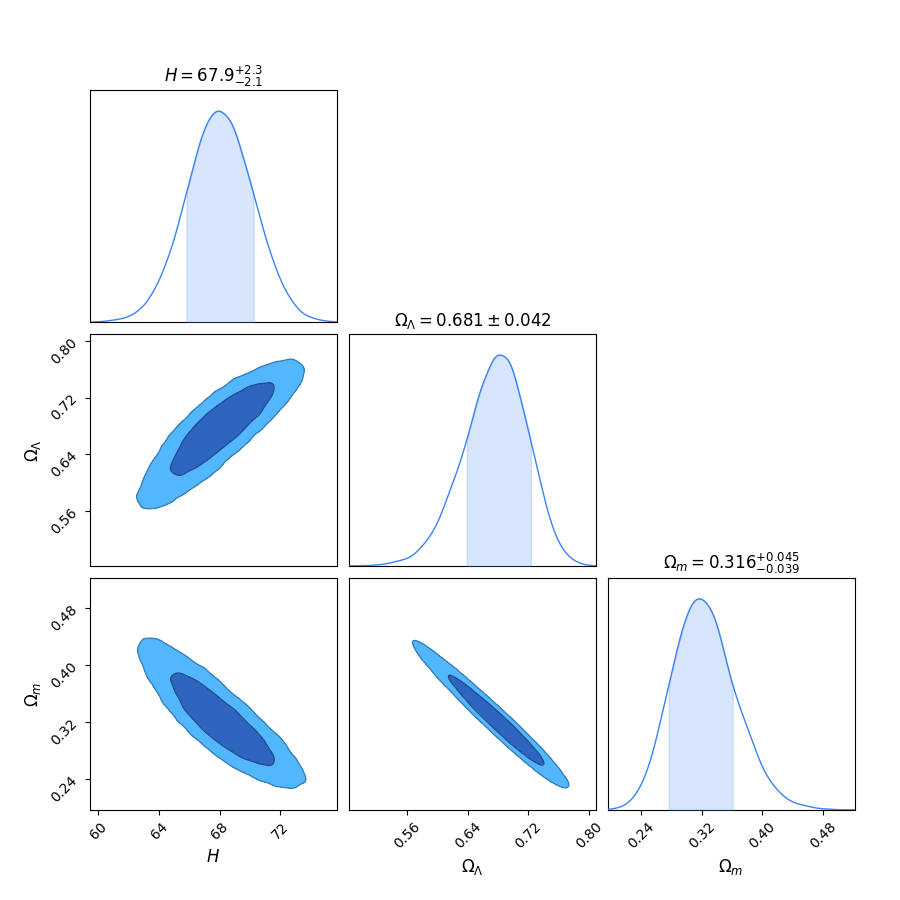}
\caption{The 68\% and 95\% confidence regions of the joint and marginal posterior probability distributions of $H$, $ \Omega_{\Lambda}$ and $\Omega_{m}$ that are estimated from parameter constraints with the data from Table. \ref{CC data}. This figure reflects the results of parameter constraints under the $\alpha = 0$ and $ \Omega_{k}=0$ conditions. The unit of $H$ is $\mathrm{km\ s^{-1}} \mathrm{Mpc^{-1}}$ and the same applies to the pictures that follow.}
\label{ms2025-0469fig1}
\end{figure}

In the initial case, we consider the case where $\omega = 0$ and $\Omega_{k} = 0$ (which means curvature $k = 0$). We then examine the constraints imposed on the set of free parameters $p = [H, \alpha, \Omega_{\Lambda}, \Omega_{m}]$, with particular emphasis on $\alpha$ and $\Omega_{m}$, as the stability of this model is determined by the relationship between these two parameters. The stability is corroborated by the findings in \cite{first_w=0} as previous research.

Next, we will discuss the details of each figure. The cosmological parameters presented in each figure are summarized in Table. \ref{Hubble Data}. In the following figures, the mean values and their $ 1\sigma$ confidence intervals for the parameter constraints are indicated above the corner plot, with the $ 1\sigma$ intervals shaded in light blue on the single-function plots. In the composite plots, the dark blue regions represent the $ 1\sigma$ confidence intervals, while the light blue regions represent the $ 2\sigma$ confidence intervals ($ 95\%c.l.$).

\begin{figure}
\centering
\includegraphics[width=\linewidth]{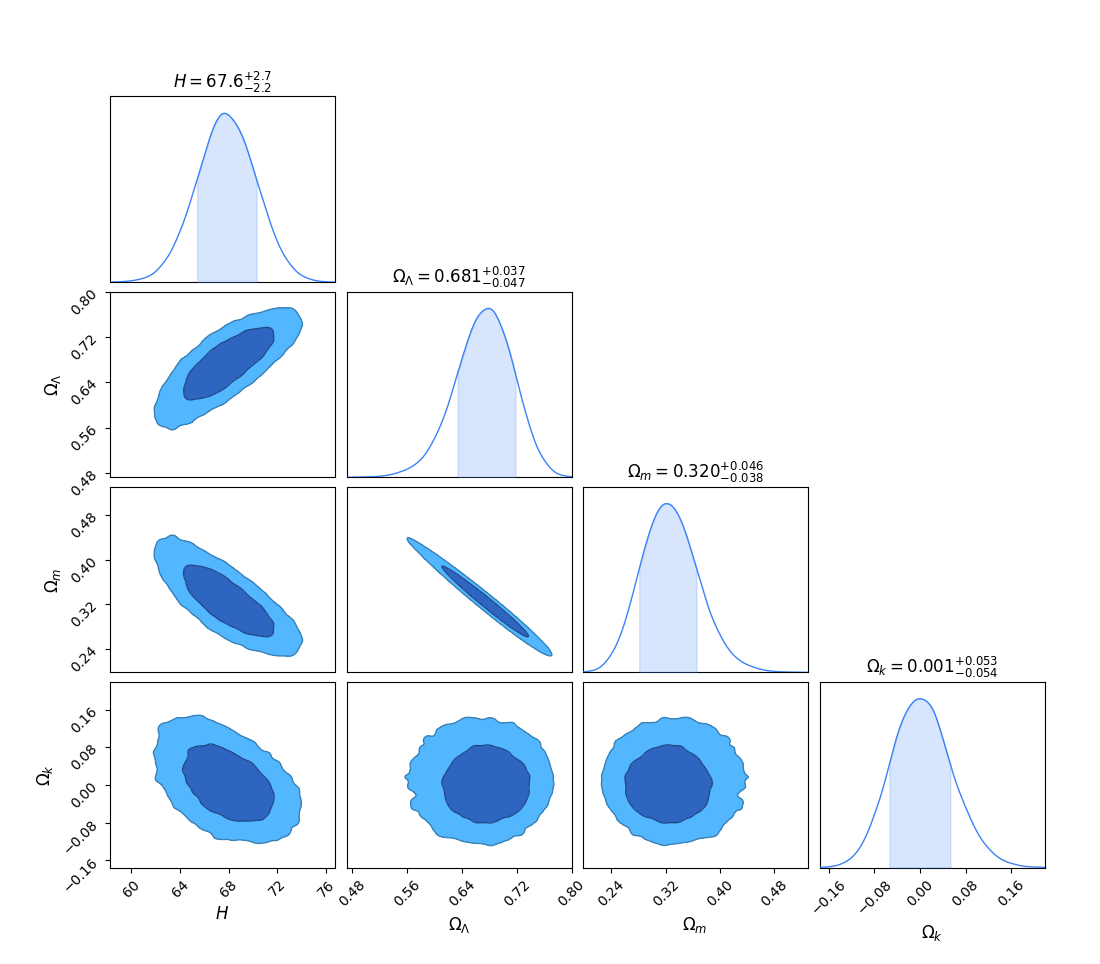}
\caption{The 68\% and 95\% confidence regions of the joint and marginal posterior probability distributions of $H$, $\Omega_{\Lambda}$, $\Omega_{m}$ and $\Omega_{k}$ that are estimated from parameter constraints with the data from Table. \ref{CC data}. This figure reflects the results of parameter constraints under the $\alpha = 0$ and $ \Omega_{k}$ can be varied conditions.}
\label{ms2025-0469fig2}
\end{figure}

Fig. \ref{ms2025-0469fig1} represents the simplest model, which does not include the curvature density parameter $\Omega_{k}$ or torsion $\alpha$. Fig. \ref{ms2025-0469fig2}, compared to Fig. \ref{ms2025-0469fig1}, only adds the $\Omega_{k}$. In the comparative analysis in the later section, these two figures will serve as the zero-control, but for now, we will refrain from further analysis or discussion of them.

\begin{figure}
\centering
\includegraphics[width=\linewidth]{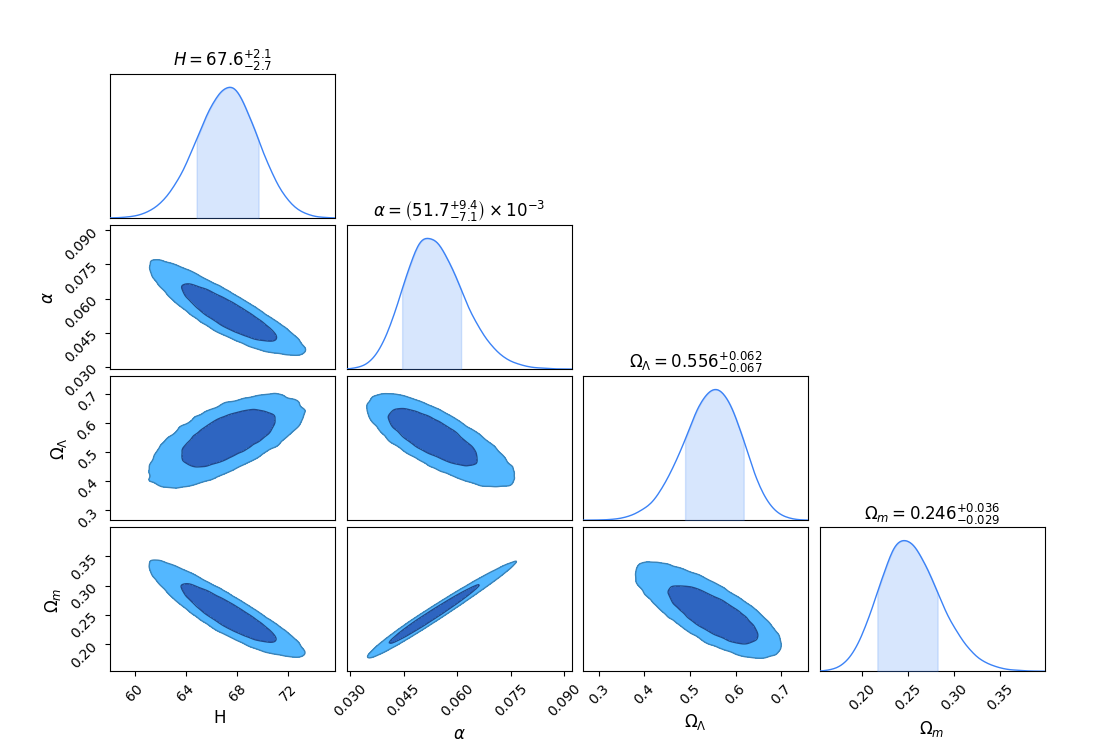}
\caption{The 68\% and 95\% confidence regions of the joint and marginal posterior probability distributions of $H$, $\alpha$, $\Omega_{\Lambda}$ and $\Omega_{m}$ that are estimated from parameter constraints with the data from Table. \ref{CC data}. This figure reflects the results of parameter constraints under the $ \Omega_{k} = 0$ and $ \alpha$ can be varied conditions with BBN limitation ($\Omega_{b}h^{2}= 0.022353\pm 0.00033 $). }
\label{ms2025-0469fig3}
\end{figure}

The results shown in Fig. \ref{ms2025-0469fig3} indicate that the constraints on all parameters approximately follow a normal distribution, with the exception of parameters $ \alpha$ and $ \Omega_{m}$, which exhibit a notable linear relationship. This linearity is attributed to the constraints on baryon density parameter $ \Omega_{b}h^{2} $ imposed by BBN. These findings suggest that the universe is stable under this torsion field model.

Furthermore, when comparing our constrained results for the Hubble constant, $ H =  67.36^{+2.1}_{-2.7} \ \mathrm{km\ s^{-1}} \mathrm{Mpc^{-1}} $, with the Planck 2018 observations $H =  67.36\pm 0.54 \ \mathrm{km\ s^{-1}} \mathrm{Mpc^{-1}} $ , we observe a close agreement between the two. This indicates that the model effectively simulates the role of various components in the evolution of the universe.

\begin{figure}
\centering
\includegraphics[width=\linewidth]{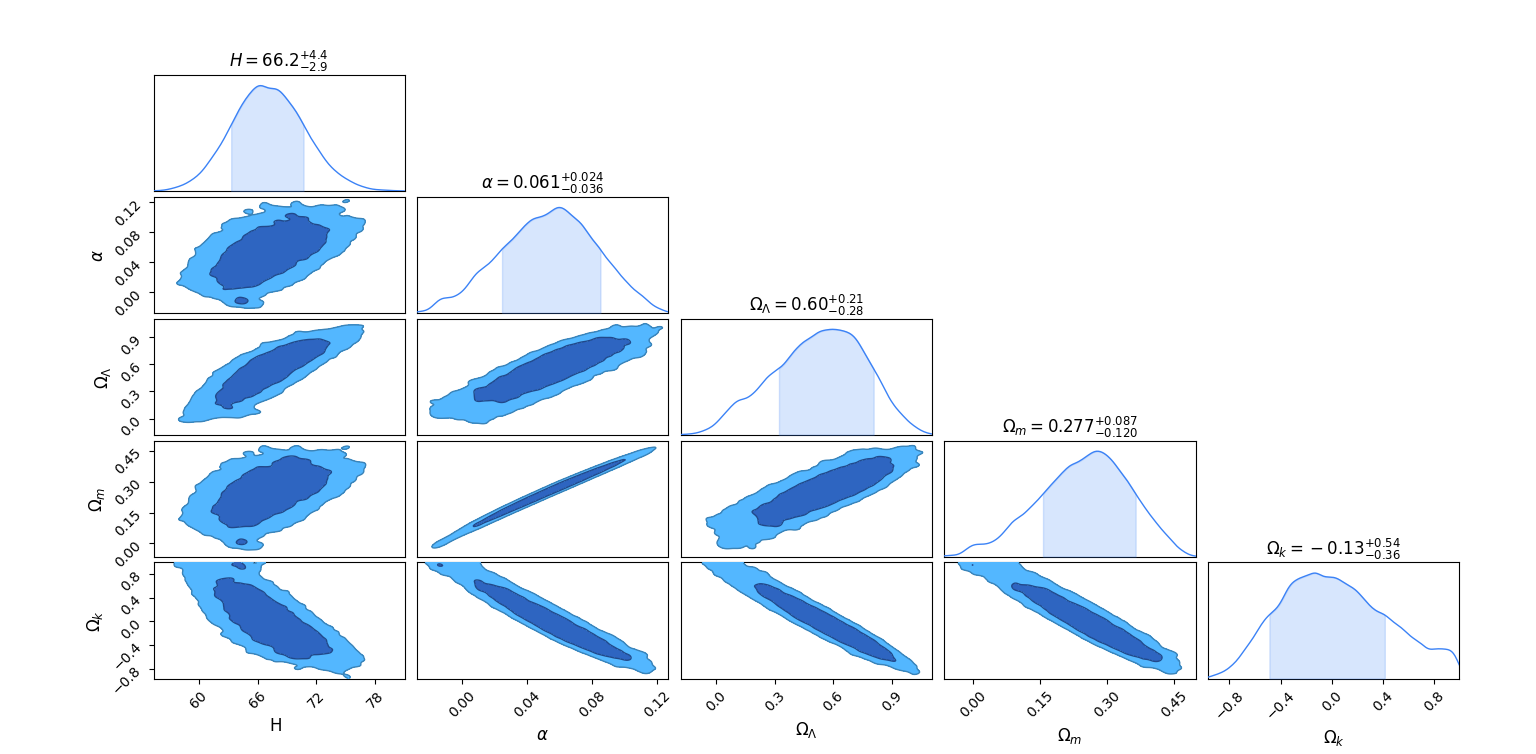}
\caption{The 68\% and 95\% confidence regions of the joint and marginal posterior probability distributions of $H$, $\alpha$, $ \Omega_{\Lambda}$, $\Omega_{m}$ and $ \Omega_{k}$ that are estimated from parameter constraints with the data from Table. \ref{CC data}. This figure reflects the results of parameter constraints under the $ \Omega_{k}$ and $ \alpha$ can be varied conditions with BBN limitation.}
\label{ms2025-0469fig4}
\end{figure}

For Fig. \ref{ms2025-0469fig4}, we introduced the curvature density parameter $\Omega_{k}$ as an additional variable, and used BBN constraints on baryon density as a limiting condition. It is easy to find that the linear relationship between $\alpha$ and $ \Omega_{m}$ nearly influences the constraint outcomes of all parameters, particularly the composite images related to $\Omega_{k}$. This suggests that under this model, the curvature density parameter is highly sensitive to variations in the torsion field.

\begin{figure}
\centering
\includegraphics[width=\linewidth]{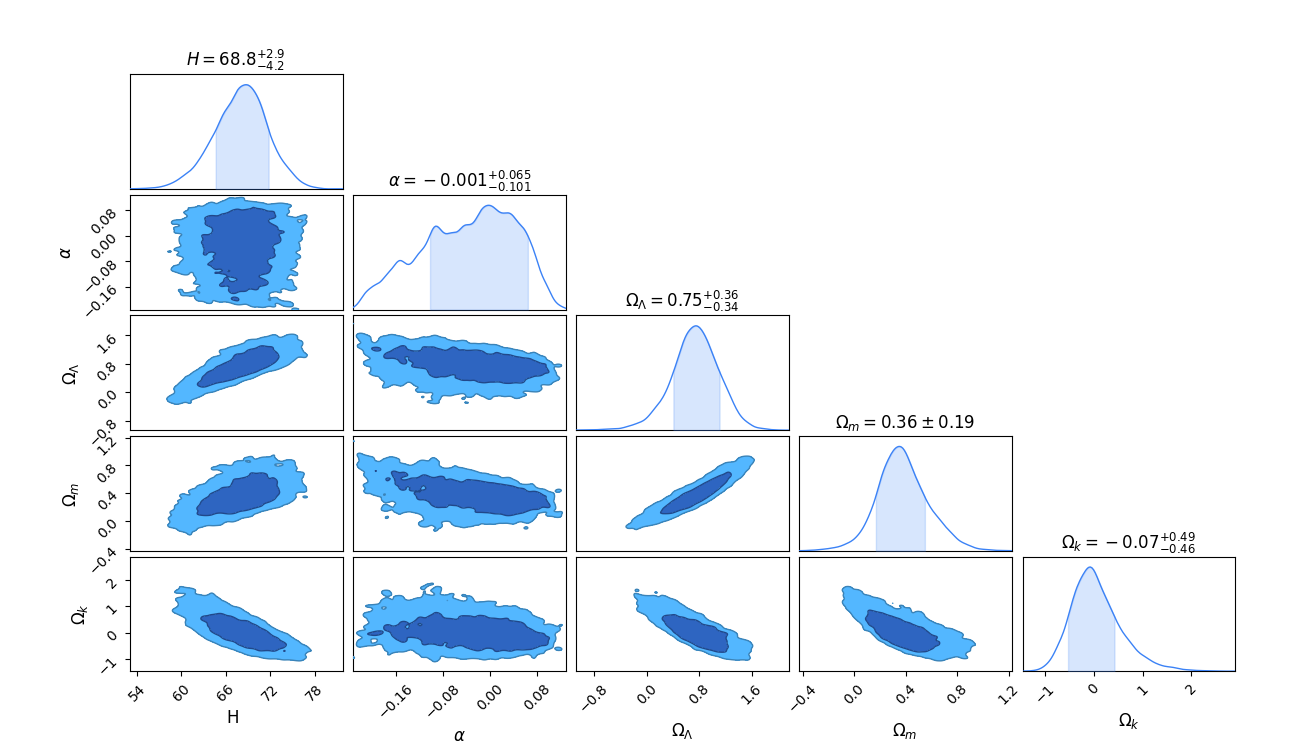}
\caption{The 68\% and 95\% confidence regions of the joint and marginal posterior probability distributions of $H$, $\alpha$, $ \Omega_{\Lambda}$, $\Omega_{m}$ and $ \Omega_{k}$ that are estimated from parameter constraints with the data from Table. \ref{CC data}. This figure reflects the results of parameter constraints under the $ \Omega_{k}$ and $ \alpha$ can be varied conditions with Planck satellite observation $\Omega_{k} = 0.0007\pm 0.0019$ limit.}
\label{ms2025-0469fig5}
\end{figure}

Next, we turn to Fig. \ref{ms2025-0469fig5}, where we utilize the Planck 2018 observations of parameter $\Omega_{k}=0.0007\pm0.0019$ as a constraint. It is evident that, apart from the composite images associated with $\Omega_{k}$, all results align consistently with Fig. \ref{ms2025-0469fig1}. Upon examining the images related to the curvature density parameter, it becomes apparent that variations in the strength of the torsion field have minimal impact on the other parameters.

This observation does not contradict the findings from Fig. \ref{ms2025-0469fig4}. Upon integrating the results of Fig. \ref{ms2025-0469fig4} and Fig. \ref{ms2025-0469fig5}, it appears that the strength of the torsion field and the curvature density parameter may have a deeper underlying connection. The uncertainties in the measurements of these two parameters may exhibit a product relationship akin to Heisenberg uncertainty, warranting further investigation.

\subsection{Results comparison and discussion}
After a series of intuitive insights into the aforementioned figures, we now return to the analysis of the physical framework. It is well known that torsion and curvature are two independent geometric parameters. In classical theory, curvature alone is sufficient to describe the distortion of spacetime, whereas in EC theory, both torsion and curvature are required for a complete description. Notably, due to the profound connection between torsion and spin in EC theory, there has been a growing interest in identifying physical phenomena or matter corresponding to torsion in actual observations, with dark matter being one of the prominent candidates.

In this paper, we propose a hypothesis that the Hubble parameter is proportional to the torsion, \(H = -\alpha\phi\), thereby linking the rate of cosmic expansion to geometric properties. While this is not the first theoretical attempt to establish such a connection (as curvature \(k\) has long been incorporated in the FLRW metric), torsion's inherent asymmetry indirectly influences the dynamical evolution of matter. This affects the large-scale structure of the universe in a manner distinct from curvature, and directly influences the rate of cosmic expansion. Further combining with the numerical simulation results in \cite{Granda_2019}, we can consider that this assumption conforms to physical logic.

To explore the relationship between various physical parameters in the figures, we examine several comparisons to identify their internal connections.

Let us first focus on the control group: Fig. \ref{ms2025-0469fig1} and Fig. \ref{ms2025-0469fig2}, to determine the effect of the presence or absence of the curvature parameter on the constraint results, without introducing torsion. It is readily apparent that the two show almost no difference, and in Fig. \ref{ms2025-0469fig2}, the joint normal distribution of \(\Omega_k\) with respect to \(\Omega_{\Lambda}\) and \(\Omega_m\) is largely uncorrelated, indicating that under general conditions, the curvature density parameter decouples from the other two.

To better assess the sensitivity of the original model to the curvature parameter and to establish a more advanced control group for subsequent analysis, we add $\Omega_{k}$ to the parameters of the control group based on the previous setup. In Fig. \ref{ms2025-0469fig1} and Fig. \ref{ms2025-0469fig3}, these figures illustrate the differences between models with and without torsion, under the BBN constraint, in the absence of curvature. When excluding the images related to \(\alpha\), slight perturbations are present, but the overall consistency remains. However, focusing on \(\alpha\), it shows a noticeable correlation with other parameters, particularly exhibiting a positive linear relationship with \(\Omega_m\). This connection likely arises because \(\alpha\), as a potential physical representation of dark matter, is part of the matter density parameter, thereby sharing the same increase-decrease relationship.

Following the aforementioned approach, naturally, we allow \(\Omega_k\) to vary in order to assess the simulation results for models with and without torsion under BBN constraints. This comparison can be observed in Figures \ref{ms2025-0469fig2} and \ref{ms2025-0469fig4}. It becomes evident that even when disregarding \(\alpha\)-related images, the results are noticeably different from the comparison between Fig. \ref{ms2025-0469fig1} and Fig. \ref{ms2025-0469fig3}. The relationship between \(\Omega_k\) and \(\Omega_{\Lambda}\) and \(\Omega_m\) shifts to a negative linear correlation, while the relationship between \(\Omega_m\), \(H\), and \(\Omega_{\Lambda}\) becomes positively correlated. This anomaly likely arises because, in previous cases, the sum of the density parameters equaled one, whereas the introduction of \(\alpha\) causes the sum to deviate from one, and this deviation is positively correlated with \(\alpha\). As a result, all density parameter relationships are altered.

Moreover, it is not only the addition of $\Omega_{k}$ that affects the parameter constraint results, variations in the constraint conditions also lead to different outcomes. We compared Fig. \ref{ms2025-0469fig4} and Fig. \ref{ms2025-0469fig5} above to explore the differences in parameter trends under BBN and Planck constraint conditions. Aside from the \(\alpha\)-related images, the two are nearly identical. However, in the case of \(\alpha\), the behavior of \(\Omega_m\) and \(\Omega_{\Lambda}\) shows opposite trends, and \(H\) exhibits almost no response to changes in \(\alpha\). This inverse relationship is likely due to the limitations of the constraint conditions themselves. For example, the BBN condition directly constrains \(\Omega_m\) and \(H\), making them dependent variables. According to the previous conclusions, \(\alpha\) shares the same increase-decrease relationship with the former, causing the sum of density parameters to increase, with \(\Omega_{\Lambda}\) also positively correlated, consistent with the graphical conclusions. However, under the Planck condition, such a relationship is not emphasized, and since the value of \(\Omega_k\) is already small, the sum of density parameters remains close to one, which narrows the reasonable range of \(\alpha\), ultimately leading the model to degenerate into a result similar to that of Fig. \ref{ms2025-0469fig1}.

\section{Conclusions}
In this paper, our primary work is to evaluate the plausibility of introducing torsion into the $\Lambda$CDM model (\citealt{Anselmi_2023}). By solving the ECKS equations, torsion is incorporated into the cosmological dynamical evolution equations, specifically the Friedmann equations, in a manner corresponding to spin. We apply Bayesian perspective supplemented with the MCMC method to constrain the parameters of this model. The following numerical conclusions can be drawn:

(1). Under the condition $\Omega_{k} = 0$ and with BBN constraints ($\Omega_{b}h^{2}= 0.022353\pm 0.00033 $), we obtain
\\ $H = 67.6^{+2.1}_{-2.7} \ \mathrm{km\ s^{-1}} \mathrm{Mpc^{-1}} $ and $\alpha = 0.052^{+0.009}_{-0.007}$;

(2). Allowing $\Omega_{k}$ to vary freely, with the same constraint in 1, we obtain $H = 66.2^{+4.4}_{-2.9} \mathrm{km\ s^{-1}} \mathrm{Mpc^{-1}} $, \\ $\alpha = 0.061^{+0.024}_{-0.036}$  and $\Omega_{k} = -0.13^{+0.54}_{-0.36}$;

(3). When changing the constraint condition to the Planck 2018 observational results ($\Omega_{k}=0.0007\pm0.0019$), we get $H = 68.8^{+2.9}_{-4.2}\ \mathrm{km\ s^{-1}} \mathrm{Mpc^{-1}} $, $\alpha = -0.001^{+0.065}_{-0.101}$, and \\ $\Omega_{k} = -0.07^{+0.49}_{-0.46}$.

It can be observed that the estimated values of the Hubble parameter and the associated standard deviations under the three conditions show very little difference. This is to be expected, as current observational results suggest that both torsion and curvature density parameters can only fluctuate around very small values. If the constraints in this study had not constrained on these two parameters, the differences in the Hubble parameter estimates would not be so small. However, the results for the parameter $\alpha$, which represents torsion, are distinct, showing correlations with the constraints themselves. Then, focusing on the curvature density parameter $\Omega_{k}$, aside from case 1 where it is directly set to 0, all other predictions for $\Omega_{k}$ yield negative values. This suggests that under this model, an open universe with a hyperbolic spatial structure is more likely to appear. In the future, based on this property, we can modify the geometric configuration of large-scale structures, obtain the optical information at each point of the structure through cosmological simulations, and compare it with actual observational data to verify whether the universe has a hyperbolic geometric structure.

Finally, let us turn the focus to the Hubble tension issue. It is evident that all the obtained constraints on the Hubble parameter lie within the $1\sigma$ deviation range of the results derived from CMB data ($67.4\pm{0.5} \ \mathrm{km\ s^{-1}} \ \mathrm{Mpc^{-1}}$), while they do not include the results obtained from direct measurements of Type Ia supernovae ($73.04\pm{1.04} \ \mathrm{km\ s^{-1}} \ \mathrm{Mpc^{-1}}$). This suggests that the theoretical framework of EC theory is more consistent with the former observational results. Furthermore, the influence of $\Omega_k$ on the Hubble tension introduces a destabilizing factor to the model, which may be attributed to the physical connection between spatial curvature and the torsion field. Taken together, we emphasize that the matter-like properties exhibited by the torsion field in hyperbolic geometry needs further study.

\begin{acknowledgements}
We sincerely appreciate the referee’s excellent feedback and valuable suggestions, which helped us to greatly improve our manuscript. This work was supported by National SKA Program of China, No.2022SKA0110202 and the China Manned Space Program with grant No.CMS-CSST-2025-A01.
\end{acknowledgements}

\begin{appendix}
\section{Christoffel Symbols}
In FLRW metric, we can write it in matrix form:
\begin{equation}
    g_{ab} = 
    \left(
    \begin{array}{cccc}
     -1&  0&  0& 0\\
      0&  \frac{a^{2}}{1-Kr^{2}}&  0& 0\\
      0&  0&  a^{2}r^{2}& 0\\
      0&  0&  0& a^{2}r^{2}sin^{2}\theta
    \end{array}
    \right)
\end{equation}

where $a = a(t)$ is  the scale factor, and K is the 3D-curvature. Then further calculates the Christoffel symbols:

\begin{equation}
\begin{aligned}
    \tilde{\Gamma}_{\ 11}^{0} &= \frac{a\dot{a}}{1-Kr^{2}}, \quad &
    \tilde{\Gamma}_{\ 22}^{0} &= a\dot{a}r^{2}, \quad &
    \tilde{\Gamma}_{\ 33}^{0} &= a\dot{a}r^{2}\sin^{2}\theta, \\
    \tilde{\Gamma}_{\ 01}^{1} &= \tilde{\Gamma}_{\ 10}^{1} = \frac{\dot{a}}{a}, \quad &
    \tilde{\Gamma}_{\ 11}^{1} &= \frac{Kr}{1-Kr^{2}}, \quad &
    \tilde{\Gamma}_{\ 22}^{1} &= -r(1-Kr^{2}), \quad &
    \tilde{\Gamma}_{\ 33}^{1} &= -r(1-Kr^{2})\sin^{2}\theta, \\
    \tilde{\Gamma}_{\ 02}^{2} &= \tilde{\Gamma}_{\ 20}^{2} = \frac{\dot{a}}{a}, \quad &
    \tilde{\Gamma}_{\ 12}^{2} &= \tilde{\Gamma}_{\ 21}^{2} = \frac{1}{r}, \quad &
    \tilde{\Gamma}_{\ 33}^{2} &= -\cos\theta \sin\theta, \\
    \tilde{\Gamma}_{\ 03}^{3} &= \tilde{\Gamma}_{\ 30}^{3} = \frac{\dot{a}}{a}, \quad &
    \tilde{\Gamma}_{\ 13}^{3} &= \tilde{\Gamma}_{\ 31}^{3} = \frac{1}{r}, \quad &
    \tilde{\Gamma}_{\ 23}^{3} &= \tilde{\Gamma}_{\ 32}^{3} = \cot\theta.
\end{aligned}
\end{equation}

Combined with Eq. \eqref{torsion scalar}, then we could derive the Christoffel symbols to this form which containing torsion part: (To be cautious that this result is calculate by $\Gamma_{\ (bc)}^{a} = \tilde{\Gamma}_{\ bc}^{a} + 2S_{(bc)}^{\quad a} $ and $\Gamma_{\ bc}^{a} = \tilde{\Gamma}_{\ (bc)}^{a} + 2S_{bc}^{\quad a} $ working together. )

\begin{equation}
\begin{aligned}
    {\Gamma}_{\ 11}^{0} &= \frac{a\dot{a} + 2\phi a^{2}}{1-Kr^{2}}, \quad &
    {\Gamma}_{\ 22}^{0} &= (a\dot{a} + 2\phi a^{2})r^{2}, \quad &
    {\Gamma}_{\ 33}^{0} &= (a\dot{a} + 2\phi a^{2})\sin^{2}\theta,&
    {\Gamma}_{\ 33}^{1} &= -r(1-Kr^{2})\sin^{2}\theta,  & \\
    {\Gamma}_{\ 01}^{1} &= \frac{\dot{a}}{a} + 2\phi, \quad &
    {\Gamma}_{\ 10}^{1} &= \frac{\dot{a}}{a}, &
    {\Gamma}_{\ 11}^{1} &= \frac{Kr}{1-Kr^{2}}, \quad &
    {\Gamma}_{\ 22}^{1} &= -r(1-Kr^{2}), \quad & \\
    {\Gamma}_{\ 02}^{2} &= \frac{\dot{a}}{a} + 2\phi, \quad &
    {\Gamma}_{\ 20}^{2} &=\frac{\dot{a}}{a}, &
    {\Gamma}_{\ 12}^{2} &= {\Gamma}_{\ 21}^{2} = \frac{1}{r}, \quad &
    {\Gamma}_{\ 33}^{2} &= -\cos\theta \sin\theta, \\
    {\Gamma}_{\ 03}^{3} &= \frac{\dot{a}}{a} + 2\phi, \quad &
    {\Gamma}_{\ 30}^{3} &= \frac{\dot{a}}{a},&
    {\Gamma}_{\ 13}^{3} &= {\Gamma}_{\ 31}^{3} = \frac{1}{r}, \quad &
    {\Gamma}_{\ 23}^{3} &= {\Gamma}_{\ 32}^{3} = \cot\theta.
\end{aligned}
\end{equation}

\end{appendix}

\bibliography{ms2025-0469citations.bib}{}
\bibliographystyle{aasjournal}

\end{document}